\documentclass[final,5p,times,twocolumn]{elsarticle}
\usepackage{amssymb,amsmath}

\begin{document}

\begin{frontmatter}

\title{On the cutoff parameter in the translation-invariant theory of the strong coupling polaron.
(response to comments [8] on the  paper V. D. Lakhno, SSC, 152, (2012), 621)}

\author[lak]{V.D.~Lakhno\corref{cor1}}
\ead{lak@impb.psn.ru}

\cortext[cor1]{Corresponding author}

\address[lak]{Institute of Mathematical Problems of Biology, Russian Academy of Sciences, Pushchino, Moscow Region, 142290, Russia}

\begin{abstract}
The paper is a reply to the arguments adduced by the authors of \cite{8} against the results obtained by the author in \cite{6,7}.  It is shown that these arguments are based on the erroneous approach made in \cite{8} to the strong coupling limit when the cutoff parameter is introduced in the theory.
\end{abstract}


\end{frontmatter}

Historically, the polaron theory was among the first to be developed
for the description of interaction between a particle and a field.
Being nonrelativistic, it did not involve the cutoff parameter which
is an inherent attribute of relativistic quantum theories for the
particle-field interaction. For this reason the polaron theory has
been a ground for testing various quantum field methods.

Among the fundamental problems which have been discussed throughout
the whole history of the polaron theory development is that of
whether the polaron is localized or delocalized in the strong
coupling limit \cite{1}-\cite{5}. In my recent papers \cite{6,7} I
have shown that the ground state of the polaron and bipolaron is
delocalized since its energy value was found to be lower than that
of the localized one. The approach used in \cite{6,7} which
eliminates the particle's coordinates through canonical
transformation and therefore is translation invariant by itself,
leads to a delocalized state for all the coupling constants
$\alpha$, including the limit case $\alpha\rightarrow\infty$.

In recent comments on these papers \cite{8} a conclusion was made about their fallacy. Here we reply to the arguments adduced in \cite{8}.

The results obtained in \cite{6,7} are based on the approach [9]
developed by Tulub for describing the strong coupling polaron.
Tulub's theory involves the function $q(1/\lambda)$ (Q in
designations used in \cite{8}) where $\lambda =4\alpha
a/3\sqrt{2\pi}$, $\alpha$ is the constant of electron-phonon
coupling, $a$ is a variational parameter. Tulub found $q(0) =5,75$.
At the same time the authors of \cite{8} showed that $q$ is a
monotonously increasing function of $\lambda$ and as
$\lambda\rightarrow\infty$ $q(1/\lambda)\rightarrow\infty$. They
considered the value $q(0)=5,75$ to be "non-physical" and used the
asymptotics $q(1/\lambda) =2\sqrt{3\lambda}$ as the basis for their
calculations of the polaron energy. As a result they obtained the
energy $E\sim \alpha^{4/3}$ which depends on $\alpha$ weaker than
the well-known strong coupling limit $E\sim \alpha^{2}$. Therefore
the values of the polaron energy in \cite{8} are greater than those
found in the strong coupling theory. So strange behavior of the
function $q(1/\lambda)$ uncharacteristic for physical systems has
not got the attention of the authors of \cite{8}.

Turning to the root of the matter, we note that Tulub's theory is a quantum field polaron theory in which the cutoff parameter arises naturally.

To demonstrate this fact let us choose the probe functions $f_k$ involved in \cite{9} in the form:

\begin{eqnarray}
\label {eq.1}
f_k=-\theta\left(k_{max}-k\right)V_k \exp(k^2/a^2),
\end{eqnarray}
where $\theta(x)=1$, $x>0$ and $\theta(x)=0$, $x<0$.
For $k_{max}=\infty$, $f_k$ are the functions chosen in \cite{9}.
With the use of (1) Tulub's function $q(1/\lambda)$ will be written as

\begin{eqnarray}
\label {eq.2}
q(1/\lambda)=\frac{2}{\sqrt{\pi}}\int^{y_{max}}_{0}\frac{e^{-y^2}(1-\Omega(y))dy}{(1/\lambda+V(y))^2+\pi
y^2e^{-2y^2}/4},\\ \nonumber
\Omega(y)=2y^2\left\{(1+2y^2)ye^{y^2}\int_{y}^{\infty}e^{-t^2}dt-y^2\right\},\\
\nonumber
V(y)=1-ye^{-y^2}\int_{0}^{y}e^{t^2}dt-ye^{y^2}\int_{y}^{\infty}e^{-t^2}dt,
\end{eqnarray}
where $y=k/a$, $y_{max}=k_{max}/a$.
In Tulub's paper the upper limit of integration in (2) was chosen to lie in the range

\begin{eqnarray}
\label {eq.3}
a<<k_{max}<<a\sqrt[4]{\lambda}
\end{eqnarray}
This choice of the cutoff parameter is caused by the fact that for
$y\rightarrow\infty$, $V(y)\rightarrow-3/4y^4$ and integration in
(2) must not involve the maximum of the integrand function occuring
at the point $y_0=\sqrt[4]{3\lambda/4}$, since for
$\alpha\rightarrow\infty$ $y_0\rightarrow\infty$ (see marginalia at
page 1833 in Tulub's paper \cite{9}). When the cutoff parameter lies
in the range (3), the value of $q$ in \cite{9} was found to be
$q=q(0)=5,75$. This leads to the polaron energy $E=-0,105\alpha ^2$.

The problem which remains to be clarified is whether condition (3)
is consistent with the limit $\alpha\rightarrow\infty$. If we admit
that it is, the quantity $a$ will be: $a\sim \alpha$ \cite{9}.

Expressions for $\alpha$ and (3) yield that for
$k_{max}\sim\alpha^p$ the value of $k_{max}$ can always be chosen so
that condition (3) be fulfilled for any $p$ lying in the range
$1<p<p_1$, where $p_1=3/2$. This proves the correctness of
inequalities (3).

For $p>p_1$ conditions (3) fail and integration in (2) involves the
maximum of the integrand lying on the infinity (as
$\alpha\rightarrow\infty$). The value of $q$ becomes equal to
$q(1/\lambda)=2\sqrt{3\lambda}$ and the polaron energy is $E\sim
\alpha^{4/3}$ as was found in \cite{8}.

Hence, in theory \cite{9} the variational estimate of the ground
state energy greatly depends on the choice of the probe function
$f_k$. It can be said that Tulub's choice of $f_k$ is the best (see
however comment \cite{10}). On the contrary, the choice of $f_k$
made in \cite{8} is the worst.

Notice that earlier the fact that the strong coupling limit depends
on the form of the relation between the cutoff parameter and the
constant of the electron-phonon interaction was pointed out in paper
by Gross \cite{3}. All the aforesaid suggests that depending on the
value of $p$, $q$ can take on any value in the range
$q\in(5,75\div\infty)$, as $\alpha\rightarrow\infty$. In the case of
$q=5,75$ the polaron value will be the lowest. Hence, presently the
translation-invariant approach developed in [9] is the best. For the
polaron (bipolaron), it yields lower variational estimates of the
ground state energy, as $\alpha\rightarrow\infty$, than theories
with spontaneously broken symmetry do (discussion of these problems
is given in \cite{11}), and the results obtained in \cite{6,7} raise
no doubt.

The work was supported by the Russian Foundation for Basic Research, project N 11-07-12054 and 10-07-00112.

\end{document}